\documentclass[a4paper]{jpconf}
\usepackage{amsmath}
\usepackage{amssymb}
\usepackage{graphicx, color}
\begin{document}
\title{Squeezed coherent states and the Morse quantum system}

\author{M Angelova$^1$\ and V Hussin$^2$}

\address{$^1$Intelligent Modelling Lab, School of Computing,
Engineering and Information Sciences, Northumbria University,
Newcastle NE2 1XE, UK}
\address{$^2$D\'epartement de Math\'ematiques et de
Statistique et Centre de Recherches Math\'{e}matiques,
Universit\'{e} de Montr\'{e}al, Montr\'{e}al, Qu\'ebec, H3C 3J7, Canada}

\ead{maia.angelova@northumbria.ac.uk, hussin@dms.umontreal.ca}

\begin{abstract}
The Morse potential quantum system is a realistic model for studying vibrations of atoms in a diatomic molecule. This system is very close to the harmonic oscillator one. We thus propose a construction of squeezed coherent states similar to the one of harmonic oscillator using ladder operators. Properties of these states are analysed with respect to the localization in position, minimal Heisenberg uncertainty relation, the statistical properties and illustrated with examples using the finite number of states in a well-known diatomic molecule.

\end{abstract}

\section{Introduction}

Coherent and squeezed states are known to be very important in many fields of physics. The coherent states were discovered in 1926 by Schr{\"{o}}dinger \cite{schro 26}, while the squeezed states were introduced by Kennard in 1927 \cite{kenn 27}. However, these works were, in the main, ignored or forgotten until the sixties, when these states became very popular and received a lot of attention with respect to both the mathematical and physical points of view. Among many interesting papers, let us mention the works of Glauber \cite{Glauber}, Klauder  \cite{Klauder, KS}, and Nieto  \cite{Nieto}. In the particular field of quantum optics, the books of  Walls and Milburn \cite{Walls}, Gazeau \cite {Gazeau} and Rand   \cite{Rand} are very good reading which also consider the applications. The study of squeezed states for systems admitting an infinite discrete spectrum, obtained as a generalisation of coherent states, has been recently the center of much attention (see, for example, \cite{Braunstein, Hillery, Bergou, Sasaki1, Alvarez}). 

In modern developments, coherent states are standardly defined by three equivalent ways: displacement operator method, ladder (annihilation) operator method and the minimum uncertainty method (for review see for example \cite{Nieto}). Starting first with the original definition for the case of the harmonic oscillator, coherent states have been generalised for other systems.  We can use, for example, the 
definition of Klauder \cite{Klauder01} saying that they are obtained as the following superposition of energy eigenstates $\{| \psi_{n}\rangle, n\in {\mathbb{N}}\}$
\begin{equation}
\psi(z) ={\frac{1}{\sqrt{{\cal N} (|z|^{2})}}}\sum_{n \in I} \frac{z^{n}}{\sqrt{\rho(n)}}| \psi_{n}\rangle.
 \label{genecs}
\end{equation}
The sum is taken over all the discrete values of $n$. The parameter $z$ is  a complex variable in general, ${\cal N}$ is a normalization factor and $\{\rho(n), n\in {\mathbb{N}}\}$ is a set of strictly positive parameters, usually depending on the energy of the system under consideration. These last quantities correspond to a moment problem (see \cite{Klauder01} for details).

Introducing ladder operators $A^-$ and  $A^+$ acting on the
energy eigenstates as
\begin{equation}
A^-| \psi_{n}\rangle=\sqrt{k(n)}\ | \psi_{n-1}\rangle, \ A^+ | \psi_{n}\rangle=\sqrt{k(n+1)}\ | \psi_{n+1}\rangle,
\label{ladder0}
\end{equation}
we can see that these coherent states are defined as eigenstates of $A^-$ and we have \cite{Klauder01}
\begin{equation}
\rho(n)=\prod_{i=1}^{n} k(i), \quad \rho(0)=1.
\label{rho}
\end{equation}
For the harmonic oscillator, we know that the usual coherent states are obtained when $\rho(n)=n!$, {\it ie} the product of its shifted energies.

Second, one way of defining squeezed coherent states is as eigenstates of the operator $A^-+\gamma A^+$, so that we get \cite{Alvarez}
\begin{equation}
\psi(z, \gamma)={\frac{1}{\sqrt{{\cal N}(z, \gamma)}}}\sum_{n \in I} \frac{Z(z, \gamma, n)}{\sqrt{\rho(n)}}| \psi_{n}\rangle,
 \label{genscs}
\end{equation}
where now $Z(z, \gamma, n)$ is interestingly a polynomial in a new complex variable obtained from the parameters $z$ (coherence) and $\gamma$ (squeezing) (see Section 2 for details).

Now, for a quantum system which admits a finite discrete spectrum like
the one which involves the Morse potential, construction of coherent
states has been adapted  \cite{Dong1, Roy, Daoud}.
In a recent paper \cite{Angelova}, we have used ladder operators
\cite{Frank, Dong2} to construct different types of coherent states of the Morse
potential and we have compared them with the so-called Gaussian coherent
states \cite{Fox}. In particular, such a construction has been
inspired by the approach mentioned above (see the formula (\ref{genecs})) but where the set $I$ of values of $n$ is now finite. The coherent states are not exactly eigenstates of the
annihilation operator $A^-$ but we have shown \cite{Angelova} that,
in practice, the last term on the right hand side of the sum in (\ref{genecs}) does not
contribute significantly. In some approaches (see, for example,
\cite{Draganescu}) these states are called pseudo-coherent states.

To our knowledge, squeezed coherent states for the Morse potential
have not been constructed. One of the aims of this paper is thus to
show that such a construction can be closely related to
the one for infinite spectrum systems. In fact, these states would be almost eigenstates of a
linear combination of the ladder operators. Another aim is to
investigate the behaviour of those squeezed coherent states with respect to localization, minimal uncertainty relation, quantum noise and also to compare their statistical properties with the ones of the harmonic oscillator system.

In Section 2 we give a review of relevant results on squeezed coherent states and minimal uncertainty relations. Starting with the definition of the Morse model and its ladder operators, we define, in Section 3, the corresponding squeezed coherent states. The uncertainty relation, localization, quantum noise and statistical properties of the squeezed states of diatomic molecules are investigated in Section 4, followed by conclusions in Section 5.
\section{General results on uncertainty relations and squeezed coherent states\label{sec:general}}

In this section we make a review of the construction of squeezed
coherent states for quantum systems admitting an infinite discrete
spectrum. These results are well-known and given in many different
contributions (see, for example \cite{Walls}). We present them in a summarized and consistent way in
order to use them in the next section.

\subsection{The quantum harmonic oscillator}

The original development  \cite{Glauber} of coherent states, in the case of the quantum harmonic oscillator, has been closely related to the fact that they minimize the Heisenberg uncertainty relation. Indeed,  a "coherent" superposition of the eigenstates produces minimum uncertainty states with respect to the position operator $x$, and the momentum operator $p$ :
\begin{equation}
{(\Delta  x)}^{2}{(\Delta  p)}^2  = \frac{\hbar^2}{4},
 \label{mur1}
\end{equation}
and, moreover, they satisfy
\begin{equation}
{(\Delta x)}^{2}= \frac{\hbar}{2 m \omega}, \  {(\Delta  p)}^2  = \frac{\hbar \ m \omega}{2}.
 \label{murexact}
\end{equation}
The mean value and dispersion of a given operator $A$ are
defined, as
usual, by
\begin{equation}
\langle  A \rangle = \bigl\langle \psi |  A | \psi \bigr\rangle, \quad (
\Delta  A)^2 =
\langle  A^2 \rangle - \langle  A \rangle^2,
\end{equation}
for a normalized state $| \psi \rangle$ describing the evolution of a quantum system.

The harmonic oscillator hamiltonian is given, as usual, by
\begin{equation}
H_{ho}=-\frac{\hbar^2}{2m} \frac{d^2}{dx^2}+\frac12 m \omega^2 x^2= \frac{1}{2m}  p^2+\frac12 m \omega^2 x^2,
\end{equation}
where
the coherent states are built as eigenstates of the annihilation operator $a^-$ :
\begin{equation}
a^-\ \psi (z, x)=z\ \psi(z, x),\quad z\in {\mathbb C}.
\label{cshoequ}
\end{equation}
They take the explicit form:
\begin{equation}
\psi (z,x)=e^{-{|z|^{2}\over{2}}}\sum_{n=0}^{\infty}{{z^{n}}\over{(n!)^{{{1}\over{2}}}}}\phi_{n}( x),
\label{csho}
\end{equation}
where the energy eigenstates of the harmonic oscillator are written as :
\begin{equation}
\phi_{n}(x)=
(\frac{m \omega}{\hbar \pi})^{\frac{1}{4}} 
\frac{1}{\sqrt{2^{n}(n!)}}
	e^{-\frac{m \omega}{2 \hbar} x^2}
	 {\rm Hermite}[n,\sqrt{\frac{m\omega}{\hbar}} x],
\label{Hermitefunctions}
\end{equation}
and ${\rm Hermite}[n,x]$ is the $n$th Hermite polynomial in $x$.
The ladder operators are defined as
\begin{equation}
a^-=\sqrt{\frac{m \omega}{2 \hbar}}(x+\frac{i}{m \omega} p), \ a^+=\sqrt{\frac{m \omega}{2 \hbar}}(x-\frac{i}{m \omega} p)\end{equation}
and their action on the eigenstates $\phi_{n}(x)$ is given by
\begin{equation}
a^- \ \phi_n=\sqrt{n} \ \phi_{n-1}, \quad a^+\phi_n=\sqrt{n+1}\ \phi_{n+1}.
\label{holadder}
\end{equation}
The harmonic oscillator hamiltonian $H_{ho}$ can be factorized as
\begin{equation}
H_{ho}=\hbar \omega (a^+ a^- +{1\over 2})= \hbar \omega (N +{1\over 2}),
\end{equation}
where $N$ is the number operator defined as $N  \phi_n=n\ \phi_n$. We have
  \begin{equation}
[H_{ho},a^+] = [N,a^+]= a^+,\
[H_{ho},a^-] = [N,a^-]= -a^-
\end{equation}
and
\begin{equation}
[a^-,a^+] = I,
\end{equation}
where $I$ is the identity operator. The set of operators $\{a^+, a^-, I\}$ generates the algebra $h(2)$.
The coherent states  (\ref{csho}) are normalizable, continuous in $z$, stable in time, and they verify the resolution of the identity operator. Moreover, they form a complete set of states but they are not orthogonal. 

They are also obtained from the action of a "displacement" operator $D(z)$ on the fundamental energy eigenstate $\phi_0$ :
\begin{equation}
\psi(z,  x)=D(z)\phi_{0}( x),
\end{equation}
where
 \begin{equation}
 D(z)= e^{ z a^+ - z^* a^-}=e^{\frac{\vert z\vert^2}{ 2}}e^{z a^+} e^{ - z^* a^-},
 \label{displacement}
\end{equation}
which satisfies $D(z)^{-1} a^- D(z)=a^-+z$.

Let us note that, in the case of coherent states, due to the relation (\ref{murexact}),
neither ${(\Delta x)}$ nor ${(\Delta p)}$ tends to zero, so even an ideal laser operating in a pure coherent state will still possess a "quantum noise" \cite{Walls}. Thus  more general classes of minimum uncertainty states
are used in quantum optics \cite{Walls} where ${(\Delta x)}$ or
${(\Delta  p)}$ could tend to zero. They are known
as squeezed states and also referred to as two-photon
coherent states \cite{Yuen}. They can be constructed as the
solutions of the eigenstate equation
\begin{equation}
(a^-+\gamma a^+)\ \psi(z,\gamma, x)=z\ \psi(z,\gamma,  x), \quad z, \gamma\in {\mathbb C},
\label{scseigenequ}
\end{equation}
which is a direct generalization of (\ref{cshoequ}). The mixing of $a^-$ and $a^+$ is said to be controlled by a squeezing parameter ($\gamma$). These states can thus be called squeezed coherent states because coherent states are special solutions when $\gamma=0$. Note that $| \gamma|<1$ for the states  to be normalizable.

As a consequence of (\ref{scseigenequ}) these states minimize the Schr\"{o}dinger-Robertson uncertainty relation \cite{Merzbacher}  which becomes the usual Heisenberg uncertainty one for $\gamma$ real. We get explicitly
 \begin{equation}\label{uncert}
{(\Delta x)}^{2}{(\Delta p)}^2  =\Delta^2=  \frac{1}{4}  \big(1+(\frac{2 (Im \gamma)}{1-| \gamma|^2})^2\big),
\end{equation}
together with
\begin{equation}
{(\Delta x)}^{2}=  \frac{1- Re \gamma}{1-| \gamma|^2}-\frac{1}{2},\ {(\Delta p)}^2  =  \frac{1+Re \gamma}{1-| \gamma|^2}-\frac{1}{2}.
\label{smur}
\end{equation}
When $Im \gamma\neq 0$ , we get  ${(\Delta x)}^{2}{(\Delta p)}^2  >\frac14$ so that this quantity never reaches its minimum value while if $Im \gamma= 0$, the dispersions in $x$ and $p$ now satisfy :
\begin{equation}
{(\Delta x)}^{2}=  \frac{1}{1+ \gamma}-\frac{1}{2},\ {(\Delta p)}^2  =  \frac{1}{1- \gamma}-\frac{1}{2},
\label{dispersionxp}
\end{equation}
making possible the reduction of the "quantum noise" on one of the 
observables with the price of increasing it on the other. In the
following, we will treat the case when the quantum noise is reduced
on the observable $x$ because we will be concerned with a
good localization in the position. In some papers,  the "total
noise" has been introduced as \cite{Hillery}:
\begin{equation}\label{noise}
T={(\Delta  x)}^{2}+{(\Delta  p)}^{2}=-1+\frac{2}{1-\gamma^2}.
\end{equation}
We thus see that this quantity reaches its minimum value only in the coherent states of the harmonic oscillator.

Note that, the eigenvalue equation (\ref{scseigenequ}) gives us, in terms of the variables $z, \ \gamma \in {\mathbb C}$,  the following solution for the harmonic oscillator squeezed coherent states \cite{Yuen, Alvarez}:
\begin{equation}
\psi_{oh}(z,\gamma, x)={\frac{1}{\sqrt{{\cal N} (z, \gamma)}}}\sum_{n=0}^{\infty}{{Z_{oh}(z,\gamma,n)}\over{(n!)^{{{1}\over{2}}}}}\phi_{n}( x),
\label{scs-ho}
\end{equation}
where
\begin{equation}
Z_{oh}(z,\gamma,n)=\sum_{i=0}^{[\frac{n}{2}]}{\frac{n!}{i! (n - 2 i)!}} (-\frac{\gamma}{2})^i {z^{(n - 2 i)}}=(\frac{\gamma}{2})^{\frac{n}{2}} {\rm Hermite}[n,{\frac{z}{\sqrt{2\gamma}}}]
\label{SCSho-coeff}
 \end{equation}
and ${\cal N} (z, \gamma)$ is a constant of normalization given by
 \begin{equation}
{\cal N} (z, \gamma)=\sum_{n=0}^{\infty}{{|Z_{oh}(z,\gamma,n)|^2}\over{n!}}.
 \end{equation}
In (\ref{SCSho-coeff}), we see that the definition of the Hermite polynomials has been extended on ${\mathbb C}$. These polynomials have been shown to have interesting properties in terms of orthogonality, measure and resolution of the identity \cite{Szafraniec}.

When $\gamma=0$, we recover the usual coherent states since $Z_{oh}(z,0,n)=z^n$. For the squeezed vacuum ($z=0$), we get
\begin{equation}
Z_{oh}(0,\gamma,2n)={\frac{(2n)!}{n!}} (-\frac{\gamma}{2})^n, \quad Z_{oh}(0,\gamma,2n+1)=0.
 \end{equation}
The associated states take the form
 \begin{equation}
\psi_{oh}(0,\gamma,  x)={\frac{1}{\sqrt{{\cal N} (0, \gamma)}}}\sum_{n=0}^{\infty}\frac{\sqrt{(2n)!}}{n!} (-\frac{\gamma}{2})^n \phi_{2n}( x),
\label{scs-hozzero}
\end{equation}
with
 \begin{equation}
{\cal N} (0, \gamma)=\sum_{n=0}^{\infty}\frac{(2n)!}{n!^2} (\frac{\gamma}{2})^{2n}.
 \end{equation}

In the case of squeezed coherent states, let us
recall that the probability distribution $P_{oh}(z,\gamma, n)$ of the
energy eigenstate $\phi_{n}( x)$ is explicitly given by
\begin{equation}
P_{oh}(z,\gamma, n)= |\bigl\langle\phi_{n}( x) | \psi (z,\gamma, x) \bigr\rangle|^2={\frac{1}{{\cal N} (z, \gamma)}}{{|Z_{oh}(z,\gamma,n)|^2}\over{n!}}.
\label{pn}
\end{equation}

 It is  well-known (see, for example, \cite{Walls}),
that this is a Poisson distribution in the special coherent case
($\gamma=0$). Among other functions, Mandel's $Q$-parameter \cite{Mandel} has been most frequently used to study the statistical properties of those states. It measures the deviation from the Poisson distribution and thus is used to distinguish quantum processes from the classical ones. Mandel's $Q$-parameter is given by,
\begin{equation}
Q(z,\gamma)=  \frac{(\Delta N)^2-\langle  N \rangle}{\langle  N \rangle},
\label{corr}
\end{equation}
where the dispersion and mean values in $N$ are computed in the squeezed coherent states (\ref{scs-ho}) and can be written as
 \begin{equation}
\langle  N \rangle=\sum_{n=0}^{\infty}n P_{oh}(z,\gamma,n), \ (\Delta N)^2=\sum_{n=0}^{\infty}n^2 P_{oh}(z,\gamma,n)-(\sum_{n=0}^{\infty}n P_{oh}(z,\gamma,n))^2.
 \end{equation}
In the case of coherent states (Poisson statistics), we know that $Q(z,0)=0$ since it corresponds to the special case where $\Delta N=\langle  N \rangle=|z|^2$. For $\gamma\neq 0$,
we can have photon bunching  ($Q(z, \gamma)>0$ or super-Poissonian statistics) or antibunching
($Q(z, \gamma)<0$ or sub-Poissonian statistics). For the squeeze vacuum ($z=0$), we have always a super-Poissonian statistics.

Finally, it is also well-known that the squeezed coherent states for the harmonic oscillator can be obtained from the action of the displacement $D(\eta)$ and squeezed $S(\chi)$ operators on the fundamental state $\phi_{0}( x)$. More precisely, we get :
 \begin{equation}
\psi_{oh}(z,\gamma,  x)=S(\chi) D(\eta)\phi_{0}( x),
\end{equation}
where
 \begin{equation}
 D(\eta)= exp( \eta a^\dagger - \eta^* a), \ S(\chi)=exp(\frac12( \chi (a^\dagger)^2 - \chi^* a^2)),
 \label{dees}
\end{equation}
and
 \begin{equation}
 \eta=\frac{z}{\sqrt{1-|\gamma|^2}} , \ \chi=\frac{\gamma} {|\gamma|} \tanh^{-1} |\gamma|.
\end{equation}

Since the set of operators $\{a^+, a^-, I\}$ generates the algebra $h(2)$, the states solving the eigenstate equation (\ref{scseigenequ}) will be referred as $h(2)$-SCS in what follows.


\subsection{General definition of squeezed coherent states for a quantum system with infinite spectrum}

By definition, the  Schr\"{o}dinger-Robertson uncertainty relation \cite{Merzbacher}  for two arbitrary hermitian operators $X$ and $P$ is given as
\begin{equation}
{(\Delta  X)}^{2}{(\Delta  P)}^2  \geq  \frac{1}{4} \bigl( {\langle  C
\rangle}^2 +
{\langle  F \rangle}^2 \bigr) \geq \frac{1}{4} {\langle  C
\rangle}^2, \label{relschro}
\end{equation}
where $C$ and $F$ are hermitian operators and defined as
\begin{equation}
  C=-i[X,  P ], \quad  F = \bigl\{  X - \langle   X \rangle  I,  P - \langle   P \rangle  I \big\} ,
 \label{com-rel}
\end{equation}
where $\{\ ,\ \}$ denotes the anti-commutator.
If $\langle  F \rangle=0$, the Schr\"{o}dinger-Robertson uncertainty relation reduces to the usual Heisenberg uncertainty relation.

A necessary and sufficient condition \cite{Merzbacher} for quantum  states $\psi_{\zeta,\lambda}$ to minimise the Schr\"{o}dinger-Robertson uncertainty relation  (\ref{relschro}) is that $\psi_{\zeta,\lambda}$  solve the eigenstate equation:
\begin{equation}
( X+  i\lambda  P )\ \psi_{\zeta,\lambda} =  \zeta \ \psi_{ \zeta,\lambda}, \zeta, \lambda\in {\mathbb C} .
\label{egalite}
\end{equation}
As a consequence we have also in these states
\begin{equation}
{(\Delta  X)}^2 = |\lambda| \Delta, \quad {(\Delta  P)}^2 =
\frac{1}{|\lambda|
}\Delta,
\end{equation}
with
\begin{equation}
\Delta = \frac{1}{2} \sqrt{{\langle  C \rangle}^2 + {\langle  F \rangle}^2}.
\label{productovarto}
\end{equation}
The states $\psi_{ \zeta,\lambda} $ satisfying \eqref{egalite} with $|\lambda|=1$
are usually
called ``coherent" while  those with $|\lambda| \ne 1$ are called ``squeezed".

Let us mention that for $Re \lambda \neq 0$, we can see that \cite{Alvarez}
\begin{equation}
\langle  F \rangle = \frac{Im \lambda}{Re  \lambda} \langle  C \rangle,
\label{moyf}
\end{equation}
when computed in the states $\psi_{ \zeta,\lambda}$ and we get
\begin{equation}
{(\Delta  X)}^2 =  \frac{ {| \lambda|}^2}{2 |Re \lambda|} \langle  C \rangle , \quad
{(\Delta  P)}^2 =  \frac{1}{2 |Re \lambda|} \langle  C \rangle.
\label{moyab}
\end{equation}

For the harmonic oscillator, the squeezed coherent states (\ref{scs-ho}) are obtained by identifying $X$ with $ x$ and $P$ with $ p$.
Indeed, the states $ \psi_{ \zeta,\lambda}$ satisfy the eigenvalue equation:
\begin{equation}
( x + i \lambda \   p ) \ \psi_{\zeta,\lambda} =  \zeta  \  \psi_{\zeta,\lambda},\ \lambda\in {\mathbb C}
\label{ecuvapro1}
\end{equation}
or,
\begin{equation}
\frac{1}{\sqrt2} \bigl[ (1 + \lambda)  a^- + (1 -\lambda)  a^+ \bigr]  \psi_{\zeta,\lambda}  = \zeta  \psi_{\zeta,\lambda},
\label{evalpaa}
\end{equation}
which is equivalent to  (\ref{scseigenequ}) if we take
\begin{equation}
\gamma=\frac{1-\lambda}{1+\lambda},\ z=\frac{\sqrt 2\ \zeta}{1+\lambda}.
\label{parameters}
\end{equation}
Since $\langle  C \rangle=1$, we recover from (\ref{moyab}) the expected dispersions (\ref{dispersionxp}) in $ x$ and $ p$.

Similarly as for the harmonic oscillator, general squeezed coherent states may be now constructed as the solutions of the eigenvalue equation:
 \begin{equation}
(A^-+\gamma A^+)\psi(z,\gamma)=z \ \psi(z,\gamma),
\label{eigenequA1}
\end{equation}
for a quantum system with an infinite discrete energy spectrum $\{| \psi_{n}\rangle, n=0,1,...\}$. The operators $A^-$ and $A^+$ are ladder operators that satisfy the relations  given in (\ref{ladder0}). The quantity $k(n)$ is not unique and can be chosen to impose additional constraints to the ladder operators. The connection with the eigenvalue equation (\ref{egalite}), is realized by 
\begin{equation}
A^-=\frac{1}{\sqrt2} (X+iP), \ A^+=\frac{1}{\sqrt2} (X-iP).
\label{ladderxp}
\end{equation}
and the identification of the parameters $(z, \gamma)$ and $(\zeta, \lambda)$ is given in (\ref{parameters}).
 
Note that introducing the number operator $N$ in the usual way:
  \begin{equation}
N \ | \psi_{n}\rangle= n\ | \psi_{n}\rangle,
   \end{equation}
 we thus get the commutators
 \begin{equation}
  [N,A^-]=-A^-, \ [N,A^+]=A^+,
   \label{intrinsicalgebra}
 \end{equation}
  \begin{equation}
   [A^-,A^+] = k(N+1)-k(N)=C(N).
   \label{intrinsicalgebra1}
 \end{equation}
 We also get, for an arbitrary function $g(N)$,
 \begin{equation}
  [g(N),A^-]=(g(N-1)-g(N))A^-, \ [g(N),A^+]=(g(N+1)-g(N))A^+.
 \end{equation}

Squeezed coherent states based on $su(2)$ or  $su(1,1)$ algebras \cite{Bergou, Sasaki1} and also direct sums of these algebras with the algebra $h(2)$ \cite {Alvarez} have been constructed using group theoretical methods and the operators displacement  $D$ and squeezing $S$ similar to the ones of the harmonic oscillator (see (\ref{displacement})). In fact for $su(2)$ or  $su(1,1)$ algebras, $k(n)$ is a quadratic function of $n$.

More generally, equation (\ref{eigenequA1}) may be solved by using a direct expansion of $\psi(z,\gamma)$ in the form
\begin{equation}
\psi(z,\gamma)={\frac{1}{\sqrt{{\cal N}_{g} (z, \gamma)}}}\sum_{n=0}^{\infty}\frac{Z(z,\gamma,n)}{\sqrt{\rho(n)}}| \psi_{n}\rangle,
\label{kfact}
 \end{equation}
 with
  \begin{equation}
{\cal N}_{g} (z, \gamma)=\sum_{n=0}^{\infty}{{|Z(z,\gamma,n)|^2}\over{\rho(n)}},
 \end{equation}
 where $\rho(n)$ has been defined in (\ref{rho}) as a function of $k(n)$.
Indeed, for the case $\gamma\neq 0$, inserting (\ref{kfact}) into  (\ref{eigenequA1}), we get  a 3-term recurrence relation
 \begin{equation}
Z(z,\gamma,n+1)- z \ Z(z,\gamma,n) +\gamma\  k(n) \ Z(z,\gamma,n-1)=0,  \ n=1,2,...
\label{recurrel}
\end{equation}
Without restriction, we take $Z(z,\gamma,0)=1$ and thus  $Z(z,\gamma,1)=z$. The resolution of such a recurrence relation is given in the Appendix where we have taken the function $k(n)=n(A-n)$ which will be used in Section 3 for the Morse system. The solution is thus shown to be related to hypergeometric functions of type $ {}_2 F_1$.
Let us mention that independently of the expression of $k(n)$, the solution $Z(z,\gamma,n)$ is a polynomial in $z$ of degree $n$. Indeed, it is easy to show that  $Z(z,\gamma,n)$ could be written as
\begin{equation}
Z(z,\gamma,n)=z^n-\sum_{l=1}^{\frac{n}{2}} c(n,l) \gamma^l z^{n-2l}.
\label{bigz}
\end{equation}

Let us end this section by mentioning that, in those
squeezed coherent states, we have now, for $\gamma$ real:
\begin{equation}
{(\Delta  X)}^2 =( \frac{1}{1+ \gamma}-\frac{1}{2}) \langle C(N) \rangle , \quad
{(\Delta  P)}^2 =(\frac{1}{1- \gamma}-\frac{1}{2})\langle  C(N) \rangle,
\label{dispersionXP}
\end{equation}
since here the operator $C(N)$ is given by
\begin{equation}
C(N)=-i[X,P]=[A^-,A^+].
\end{equation}

The operators $X$ and $P$ are not the position and momentum operators except  for the case of the harmonic oscillator. In general, they involve those operators in a complicated way.

\section{The Morse potential and different types of squeezed coherent states\label{sec:ho}}

As mentioned in the introduction, the Morse potential system constitutes a better approximation of vibrations of atoms in a diatomic molecule. The new system is still very close to the harmonic oscillator. Thus, the squeezed coherent states will be constructed following the procedure given for the harmonic oscillator, except that we will deal with a finite number of eigenstates. We will show that the states constructed in this way will thus be well localized for some values of the coherent and squeezing parameters.

\subsection{The model}

The one-dimensional Morse model is given by the energy eigenvalue equation (see, for example, \cite{Frank}) 
\begin{equation}
{\cal H}\ \psi(x) =(\frac{{ p}^2}{2m_r}+V_M(x))\psi(x)= E\psi(x),
 \end {equation}
where $m_r$ is the reduced mass of the oscillating system composed of two
atoms of masses $m_1$ and $m_2$, {\it i.e.}
$\frac{1}{m_r}=\frac{1}{m_1}+\frac{1}{m_2}$. The potential is
$V_M(x)=V_0(e^{-2\beta x}-2e^{-\beta x})$, where  the space variable $x$ represents the displacement of the two atoms
from their equilibrium positions,  $V_0$ is a scaling energy
constant representing the depth of the potential well at equilibrium
$x=0$ and $\beta$ is the parameter of the model (related to the
characteristics of the well, such as its depth and width).

The finite discrete spectrum is known as
\begin{equation}
E_n=-\frac{ \hbar^2}{2m_r} \beta^2 \ {\epsilon_n}^2,
\label{energies}
\end{equation}
where
\begin{equation}
\epsilon_n=\frac{\nu-1}{2}-n=p-n, \  \nu=\sqrt{\frac{8{m_r}V_0}{ \hbar^2 \beta^2}}
\label{epsilonnu}
\end{equation}
and $\{n=0,1,2,...,[p]\}$,  with $[p]$ the integer part of
$p=\frac{\nu-1}{2}$. We see that for the Morse oscillator the consecutive energies are
not equally spaced.
The following shifted energies
\begin{equation}
e(n)=\frac {2m_r}{ \hbar^2 \beta^2}
(E_n-E_0)=\epsilon_0^2-\epsilon_n^2=n(2p-n)
\label{shifteden}
\end{equation}
are useful for the construction of squeezed coherent states.
Using the change of variable
\begin{equation}
y=\nu e^{-\beta x},
\end{equation}
we get the energy eigenfunctions, for the discrete spectrum, in terms of associated Laguerre polynomials, denoted by $L_{n}^{2\epsilon_n}$, as
\begin{equation}
\psi_n^\nu (x)= {\cal N}_n \ e^{-\frac{y}{2}} y^{\epsilon_n} L_{n}^{2\epsilon_n}(y),
\label{eigenfunMorse}
\end{equation}
where ${\cal N}_n$ is a normalization factor given by
\begin{equation}
{\cal N}_n= \sqrt{ \frac{\beta(\nu-2n-1)\Gamma(n+1)}{\Gamma(\nu-n)}}= \sqrt{ \frac{2\beta(p-n)\Gamma(n+1)}{\Gamma(2p-n+1)}}.
\label{normM}
\end{equation}
Let us mention that $p=\frac{\nu-1}{2}$ is related to the physical parameters of the Morse system. This means that it is not an integer in practice and ${\cal N}$ is never zero as expected. But mathematically speaking, if $p$ is an integer, the last normalised state is $\psi_p^\nu (x)=e^{-\frac{y}{2}} L_p(y)$ where $L_p(y)$ is the usual Laguerre polynomial.

Indeed, the orthogonality relation on the energy eigenfunctions $\psi_n^\nu(x)$ depending on the original space variable $x$ thus writes
\begin{equation}
\int_{-\infty}^\infty \psi_n^\nu (x) \psi_m^\nu (x) dx=\delta_{nm}.
\label{inner}
\end{equation}
For many applications,
it is convenient to introduce the number operator $N$ such that
\begin{equation}
N\psi_n^\nu (x)= n \ \psi_n^\nu (x).
\end{equation}
We thus see from (\ref{energies}) that the Hamiltonian operator can be related to $N$ through
\begin{equation}
{\cal
 H}\ =-\frac{ \hbar^2}{2m_r} \beta^2 \ (p-N)^2 .
 \end{equation}
\subsection{Ladder operators}

We use the definition (\ref{ladder0}) for the ladder operators of the Morse system where the set of eigenfunctions $ \{| \psi_{n}\rangle\}$ is finite and given by $\{\psi_n^\nu(x)\}$ as in (\ref{eigenfunMorse}). As mentioned in the introduction, the quantity $k(n)$ is not unique and some choices have been considered in the preceding study of coherent states for such a system \cite{Angelova}.

Here, we will consider two different types of ladder operators. The
first type is called "oscillator-like" with $k(n)=n$ since it is associated to the $h(2)$ algebra. The second type is called "energy-like"
where $k(n)=e(n)$ as given in (\ref{shifteden}). It is associated with a $su(1,1)$ algebra.
In what follows, we will use subscripts {\it o} and
{\it e} to denote an "oscillator-like" or "energy-like" type, respectively.
Let us mention that, from the commutators (\ref{intrinsicalgebra1}) acting on the
set of energy eigenstates $\{\psi_n^\nu(x)\}$, we get, in this last case:
\begin{equation}
C(n)=e(n+1)-e(n)=2(p-n-\frac12),
\label{efn}
\end{equation}
a quantity which is always positive for  $n=0,1,2,...,[p]-1$.

Though our future calculations do not need the
explicit form of the ladder operators,  we give them for
completeness. Ladder operators for the Morse potential have been obtained in
different papers  \cite{Daoud, Frank, Singh,
Sasaki}. For example, we get \cite{Frank}:
 \begin{eqnarray}
 A^-&=&-[\frac{d}{dy}(\nu-2N)-\frac{(\nu-2N-1)(\nu-2N)}{2 y}+\frac{\nu}{2}]\sqrt {K(N)}, \label{laddergen}\\
A^+&=& (\sqrt {K(N)})^{-1}[\frac{d}{dy}(\nu-2N-2)+\frac{(\nu-2N-1)(\nu-2N-2)}{2 y}-\frac{\nu}{2}] ,
\label{laddergen1}
   \end{eqnarray}
where $K(n)$ is related to $k(n)$ by
   \begin{equation}
   k(n)=\frac{n(\nu-n)(\nu-2n-1)}{\nu-2n+1}K(n).
   \end{equation}
These relations are  valid for any integer $n$ in the interval $[0,[p]-1]$. Note that, for $n=[p]$, we get an admissible energy eigenstate $\psi_{[p]}^\nu(x)$
of the Morse potential but the action of the creation operators on this state does not give zero in general.
It gives a state which may not be  normalizable with respect to our scalar product.
This problem has been already mentioned in some contributions (see, for example, \cite{Frank, Singh}).
For arbitrary $p$, the special choice
\begin{equation}
k(n)=n([p]+1-n),
\label{kspecial}
\end{equation}
leads to $A_+ \psi_{[p]}^\nu(x)=0$ and  $A_+ \psi_{[p]-1}^\nu(x)=\sqrt{[p]}\psi_{[p]}^\nu(x)$. It is not the case that we are considering in what follows but it is similar to the energy-like case.

The "oscillator-like" ladder operators are now acting on the eigenfunctions (\ref{eigenfunMorse}) of the
Morse potential as
 \begin{equation}
a^- \psi_n^\nu(x)=\sqrt{n} \ \psi_{n-1}^\nu(x), \quad a^+ \psi_n^\nu(x)=\sqrt{n+1}\ \psi_{n+1}^\nu(x),
\end{equation}
since $k(n)=n$.
We have identified
$A^-$ with $a^- $ and $ A^+$ with $a^+$ by taking
\begin{equation}
K_{o}(n)=\frac{\nu-2n+1}{(\nu-n)(\nu-2n-1)}.
\label{knoh}
\end{equation}

The "energy-like" ladder operators are obtained by identifying
$A^\pm$ with the operators denoted by $ J^\pm$ taking
\begin{equation}
K_{e}(n)=\frac{(\nu-1-n)(\nu-2n+1)}{(\nu-n)(\nu-2n-1)}= K_{o}(n) (\nu-1-n).
\end{equation}
Let us mention that we can in fact relate $a^\pm$ and $J^\pm$. Indeed, we have:
\begin{equation}
J^- =a^- \sqrt{2p-N}, \ J^+ = \sqrt{2p-N} a^+.
\end{equation}
Since now $k(n)=n(2p-n)$, we get explicitly
 \begin{equation}
J^- \psi_n^\nu(x)=\sqrt{n(2p-n)} \ \psi_{n-1}^\nu(x), \quad J^+ \psi_n^\nu(x)=\sqrt{(n+1)(2p-n-1)}\ \psi_{n+1}^\nu(x).
\end{equation}
From the expression of $C(n)$ in  (\ref{efn}), we get the following commutators (acting on the finite set of energy eigenstates $\{\psi_n^\nu, n=0,...,[p]-1]$):
\begin{equation}
[J^{\pm}, J^0 ] =\pm J^{\pm}, \ [J^+, J^- ] =-2(p-N-\frac12) =-2 J^0.
\end{equation}
The set $\{J^-, \ J^+, \ J^0=p-N-\frac12 \}$ thus generates an
$su(1,1)$ algebra.
Moreover, the energy operator may be written as
 \begin{equation}
{E_{op}}=(\frac{ \hbar^2}{2m_r} \beta^2)\  (J^+ J^- +p^2)=-(\frac{ \hbar^2}{2m_r} \beta^2)\  (J^0+\frac12)^2.
\label{energyfactor}
\end{equation}
Let us mention the case when $k(n)=(\ref{kspecial})$ is similar to the preceding choice since $k(n)$ is quadratic in $n$ but we don't have a factorisation of the energy operator ${E_{op}}$.

 \subsection{The harmonic oscillator limit}
 
Let us here recall how we get  the harmonic oscillator limit \cite{Dong2}. First, we have to shift the Morse potential $V_M$ so that it is equal to zero at the origin. We thus take
\begin{equation}
V_1=V_0(1-e^{-\beta x})^2=V_M+V_0.
\label{Morse1}
\end{equation}
and the limit is performed by choosing  $V_0= \frac{k}{2 \beta^2}$ and taking $\beta \to 0$ so that
$V_1 \to V_{OH}=\frac12 k x^2$. Note that the new Hamiltonian with potential $V_1$ has thus the energy levels shifted and we get
\begin{equation}
E_n^1=- \frac{ \hbar^2}{2m_r} \beta^2 [(\frac{\nu-1}{2}-n)^2- (\frac{\nu}{2})^2].
\end{equation}
Since, $\nu$ is given by (\ref{epsilonnu}), we get here
\begin{equation}
\nu=\frac{2 \sqrt{{m_r}k}}{\beta^2 \hbar}.
\label{nulimit}
\end{equation}
The oscillator limit is obtained when $\nu \to \infty$ giving, as expected, an infinite spectrum and the good limit for the energies
\begin{equation}
\lim_{\nu \to \infty} E_n^1= \hbar \sqrt{\frac{k}{m_r}}(n+\frac12).\nonumber
\end{equation}
Second, we have to take the limit on the ladder operators. We replace $\beta$ by its expression in terms of $\nu$ as in (\ref{nulimit}) and define $c=\sqrt{\frac{4m_r k}{\hbar^2}}$. The annihilation operator $A^-$, given in (\ref{laddergen}), thus takes the form:
\begin{equation}
A^-=  \sqrt {K(n)}[\frac{e^{\sqrt{\frac{c}{\nu}} x}}{\sqrt{c\ \nu}}(\nu-2n) \frac{d}{dx}+ \frac{e^{\sqrt{\frac{c}{\nu}} x}}{2 \nu} (\nu-2n-1)(\nu-2n)-\frac{\nu}{2}].
\end{equation}
Since $K(n)$ depends also on $\nu$, we have to take the limit carefully. Let us look at the following
limit (the coefficient of $\frac{d}{dx}$ ):
\begin{equation}
\lim_{\nu \to \infty} \sqrt {K(n)}\frac{e^{\sqrt{\frac{c}{\nu}} x}}{\sqrt{c\ \nu}}(\nu-2n)= \lim_{\nu \to \infty}(1+ \sqrt{\frac{c}{\nu}} x)\frac{\nu-2n}{{\sqrt{c\ \nu}}}\sqrt {K(n)}= \lim_{\nu \to \infty}(\frac{\nu-2n}{{\sqrt{c\ \nu}}}+ \frac{\nu-2n}{\nu} x) \sqrt {K(n)}.
\end{equation}
It means that $K(n)$ must behave as $\frac{1}{\nu}$ which is exactly what we get taking it as in (\ref{knoh}). We have now to check if we get the right limit for the other term of $A^-$. We have
\begin{equation}
\lim_{\nu \to \infty} \sqrt {K(n)} [ \frac{e^{\sqrt{\frac{c}{\nu}} x}}{2 \nu} (\nu-2n-1)(\nu-2n)-\frac{\nu}{2}]= \sqrt {K(n)}\lim_{\nu \to \infty}  [ \frac{(1+ \sqrt{\frac{c}{\nu}} x)}{2 \nu} (\nu-2n-1)(\nu-2n)-\frac{\nu}{2}].
\end{equation}
We finally find
\begin{equation}
\lim_{\nu \to \infty}  A^-= \frac{1}{\sqrt c}(\frac{d}{dx}+\frac{c}{2}x).
\end{equation}
A similar calculation gives the expected limit for $A^+$.

\subsection{Squeezed coherent states }

We are now ready to adapt the discussion, given in Section 2, in order to construct  the squeezed coherent states of the Morse Hamiltonian. Indeed, we define them as the finite sum
\begin{equation}
\Psi^{\nu}(z, \gamma, x)=  \frac{1}{\sqrt{{\cal{N}}^{\nu} (z,\gamma)}}
\sum_{n=0}^{[p]-1} \frac{Z(z,\gamma,n)}{\sqrt{\rho(n)}} \psi_n^{\nu} (x),
\label{scsMorse}
\end{equation}
where $\rho(n)$ is given in (\ref{rho}), $Z(z,\gamma,n)$ satisfies (\ref{recurrel}) and
 \begin{equation}
{\cal N}^{\nu} (z, \gamma)=\sum_{n=0}^{[p]-1}{{|Z(z,\gamma,n)|^2}\over{\rho(n)}}.
 \end{equation}

Such a definition is relevant since we have seen in the preceding subsection that the "oscillator-like" ladder operators tend to the ones of the harmonic oscillator when $k(n)=n$ and the appropriate limit is taken. Moreover, these states are  "almost" eigenstates of a linear combination of the generic ladder operators $A^-$ and $A^+$ which can be written as:
\begin{equation}
(A^-+\gamma\ A^+) \ \Psi^{\nu}(z, \gamma, x)\approx z \Psi^{\nu}(z, \gamma, x).
\end{equation}
In fact, the correction can be computed using the recurrence relation (\ref{recurrencef}) and we find
\begin{equation}
\chi^{\nu} (z,\gamma,[p],x)=\Lambda_{1}(z,\gamma,[p])\psi_{[p]-1}^{\nu} (x)+\Lambda_{0}(z,\gamma,[p]) \psi_{[p]}^{\nu} (x),
\end{equation}
where
\begin{eqnarray}
\Lambda_{1}(z,\gamma,[p])&=&\frac{1}{{\sqrt{\rho_{[p]-1}}}} Z(z,\gamma,[p]), \nonumber \\
\Lambda_{0}(z,\gamma,[p])&=&\frac{1}{{\sqrt{\rho_{[p]}}}}\gamma  k([p]) Z(z,\gamma,[p]-1).
\end{eqnarray}
In practice, the last two terms of the sum in (\ref{scsMorse}) have a very weak contribution which justifies thus the term  "almost" eigenstates used above.

Other constructions of squeezed coherent states have been considered (see, for example, \cite{Daoud, Dong2}). They implicitly use the displacement operator $D$ given in (\ref{displacement}). It must be questioned first because we are dealing with a finite number of eigenstates in (\ref{scsMorse}) (see our comments in the conclusion). Second, only one parameter is involved in this displacement operator, that is the reason why they are called coherent states by these authors. They are, in fact, special cases of our squeezed coherent states where $z$ and $\gamma$ are not independent ($\gamma\neq 0$). 

In particular, in the so-called oscillator-like squeezed coherent states, we get
 \begin{equation}
P_o(z,\gamma, n)= {\frac{1}{{\cal N}_0^\nu (z, \gamma)}}({\frac{|\gamma|}{2}})^n {{{|\rm Hermite}[n,\frac{z}{\sqrt{2\gamma}}]|^2}\over{ n!}}
\label{probaoh}
\end{equation}
with
\begin{equation}
{\cal N}_o^\nu (z, \gamma)=\sum_{n=0}^{[p]-1} ({\frac{|\gamma|}{2}})^n {{{|\rm Hermite}[n,\frac{z}{\sqrt{2\gamma}}]|^2}\over{n!}}.
\end{equation}
The mean value and dispersion of the number operator $N$ are now given by
 \begin{equation}
\langle  N \rangle_o=\sum_{n=0}^{[p]-1}n \ P_o(z,\gamma,n), \ (\Delta N)^2_o=\sum_{n=0}^{[p]-1}n^2 P_o(z,\gamma,n)-(\sum_{n=0}^{[p]-1}n\ P_o(z,\gamma,n))^2.
\label{enoh}
 \end{equation}
Note that the statistical properties of these states are thus similar to the ones of the harmonic
 oscillator since we get essentially the same quantity for  the Mandel's $Q$-parameter given by (\ref{corr}), except that the sums are now finite in $\langle  N \rangle_o$ and $(\Delta N)^2_o$.

 For the second set of states, the so-called
energy-like squeezed coherent states, we have $k(n)=e(n)=n(2p-n)$ and the recurrence relation
(\ref{recurrencef}) has been solved in terms of hypergeometric functions (see the Appendix when $A=2p$). We thus write 
\begin{equation}
Z(z,\gamma,n)=(-1)^n \gamma^{\frac{n}{2}} \frac{\Gamma(2p)}{\Gamma(2p-n)}\ {}_2 F_1\left(\begin{matrix}-n, - {\frac{z}{2 \sqrt\gamma}}+{\frac{1-2p}{2}}  \\ 1 - 2p\end{matrix};2\right), \ n=1,2,...,[p]-1.
\end{equation}

Now the probability distribution, denoted by $P(z,\gamma,n)_e$, is given by
\begin{equation}
P_e(z,\gamma, n)= {\frac{1}{{\cal N}_e^\nu (z, \gamma)}}\frac {\Gamma(2p-n) }{\Gamma(2p)n!}|Z(z,\gamma,n)|^2,
\label{pn}
\end{equation}
where
\begin{equation}
{\cal N}_e^\nu (z, \gamma)=\sum_{n=0}^{[p]-1} \frac {\Gamma(2p-n)}{\Gamma(2p)n!}|Z(z,\gamma,n)|^2.
\end{equation}
Similar expressions for $\langle  N \rangle_e$ and $(\Delta N)^2_e$ are obtained as in (\ref{enoh}).

From the construction of the ladder operators $A^-$ and $A^+$ of the Morse potential (see, (\ref{laddergen}) and (\ref{laddergen1})), we can define
the self adjoint operators $X$ and $P$ as (see (\ref{ladderxp}))
\begin{equation}
X=\frac{1}{\sqrt 2}(A^++A^-),\quad P=\frac{i}{\sqrt 2}(A^+-A^-).
\end{equation}
We easily see that $X$ and $P$ are not related to the physical observables position $x$ and momentum $p$, but we get minimal  uncertainty relation and the dispersions are given by (\ref{moyab}) with
\begin{equation}
\langle C(N) \rangle= \frac{1}{{\cal{N}}^{\nu} (z,\gamma)}
\sum_{n=0}^{[p]-1} C(n)\frac{|Z(z,\gamma,n)|^2}{\rho(n)}.
\end{equation}
For the $h(2)$-SCS, we have $\langle  C(N) \rangle_o=1$ as expected while for the $su(1,1)$-SCS, $C(n)$ is given by (\ref{efn}) and we thus get
\begin{equation}
\langle  C(N) \rangle_e = 2p-1-2\langle  N \rangle_e.
\end{equation}

Since our squeezed coherent states (in particular, the oscillator-like states) are closely related to the ones of the harmonic oscillator, we are interested in the behaviour of our states in the physical observables position $x$ and momentum $p$. In order to check if the minimal uncertainty relation is satisfied for these observables for some values of $z$ and $\gamma$, it is necessary to compute the corresponding dispersions:
    \begin{equation}\label{delx}
     (\Delta x)^2= \int_{-\infty}^\infty ( \Psi^{\nu}(z, \gamma, x)^2 x^2 dx -(\int_{-\infty}^\infty ( \Psi^{\nu}(z, \gamma, x)^2 x \ dx)^2
\end{equation}
and
 \begin{equation}\label{delp}
      (\Delta p)^2=-\int_{-\infty}^\infty  \Psi^{\nu}(z, \gamma, x) \frac{d^2\Psi^{\nu}(z, \gamma, x)}{d^2x} dx
       -( \int_{-\infty}^\infty  \Psi^{\nu}(z, \gamma, x) \frac{d \Psi^{\nu}(z, \gamma, x}{dx} dx)^2.
\end{equation}
It is done in practice by numerical integration because the functions under the integral sign are
rapidly decreasing to zero.

Moreover, we will show in the next section that, with good choices of the parameters $z$ and $\gamma$, those states are well localized with respect to the position $x$. In fact, such choices will lead to a minimization of the Heisenberg uncertainty relation close to the one obtained for the harmonic oscillator.

Finally, let us mention that time evolution for  our squeezed coherent states is computed as
usual and we get
 \begin{equation}
\Psi^{\nu}(z, \gamma, x;t)=  \frac{1}{\sqrt{{\cal{N}}^{\nu} (z, \gamma)}}
\sum_{n=0}^{[p]-1}  \frac{Z(z,\gamma,n)}{\sqrt{\rho(n)}} e^{- \frac{ i E_n}{\hbar}t} \psi_n^{\nu} (x).
\end{equation}

\section{Uncertainty, localization, quantum noise and statistical properties  of the states of diatomic molecules}

The vibrational modes of  most  diatomic molecules can
be well described using the Morse potential. In such cases, the
value of $\nu$ can be calculated from (\ref{epsilonnu}) with
published values of $m_r$, $\beta$ and $V_0$, or as most often in practice, using the ratio
between the experimentally measured molecular
harmonicity $\omega_e$ and anharmonicity $\omega_e x_e$ constants 
(see for example \cite{Herzberg50,CRC}), $\nu = {{\omega}_e/{{\omega}_e x_e}}$.

The values $\nu$ and $[p]$ have been calculated in \cite{AF07} for
many diatomic molecules. For the case of hydrogene chloride,
$^1$H$^{35}$Cl, for the ground state, $X^1 \Sigma^+$ we have $\nu\approx57.44$ and $[p]=28$. In what follows, we
will use this molecule (as in  our previous paper \cite{Angelova})
to illustrate, analyze and compare the behaviour of different types
of squeezed coherent states.

As mentioned before our construction of squeezed coherent states for the Morse potential does not give rise a priori to a minimum uncertainty relation with respect to the observables position and momentum. We thus ask the following question: what values of the parameters $z$ and $\gamma$ lead to a good localization of our states  in the position and give rise to an uncertainty relation close to the minimum? In order to help us answering this question we have summarized in the Tables 1 and 2 some results obtained from the computation of mean values of $x$, $p$ and $N$  for the two types of squeezed coherent states that have been constructed. 

Table1 is for  oscillator-like squeezed coherent states and Table \ref{table2} is for energy-like squeezed coherent states. The first column of the tables gives the values of the coherence parameter $z$, the second column the values of the squeezing parameter $\gamma$. 
As usual, coherent states are
obtained when $\gamma=0$. The dispersion in position $(\Delta
x)^2$, calculated by numerical integration of (\ref{delx}), is given
in the third column. The fourth column
represents the uncertainty $(\Delta)^2=(\ref{uncert})$ calculated as a product of
the dispersion in $x$ and $p$  . The values of the total
quantum noise $T=(\ref{noise})$, the sum of dispersion in $x$ and in $p$,  is in the fifth column. We have not given the values of $(\Delta
p)^2$ since they are easily derived from the values of $T$ and $(\Delta
x)^2$, and also because we have in mind to minimize the dispersion on $x$.
The Mandel's $Q$-parameter (\ref{corr}) is given in the last column of the tables. The chosen values for $z$ and $\gamma$ are representatives of the different behaviours of our states. We have chosen to keep at most four decimal digits for the different quantities in the Tables.

\begin{table}
\caption{Values of different observables in the oscillator-like squeezed coherent states }\label{table 1}
\centering
\begin{tabular}{l|*{6}{| c}}
 z&  $\gamma$&$(\Delta x)^2$& $(\Delta)^2$&T &$Q(z,\gamma)$\\
\hline
 0.1&0&0.0188     & 0.2526 &13.41&0  \\
  0.1&0.3&0.0115   & 0.2680 &23.30&1.1041   \\
 0.1&0.5&0.0118    &0.4114  & 34.82&1.6359  \\
 \hline
 0.3&0&0.0210 &0.2534  & 12.06 &0 \\
  0.3&0.3&0.0123   &0.2652  & 21.48&0.6170  \\
 0.3&0.5&0.0122   &0.3975  &32.47&1.4166  \\
 \hline
 0.6&0&0.0249     &0.2549 &10.26&0  \\
 0.6&0.3&0.0138   &0.2617 &18.98&0.0646  \\
 0.6&0.5&0.0130  &0.3804  &29.18&0.9099  \\
 \hline
 1&0& 0.0318&0.2604  & 8.2&0 \\
  1&0.3& 0.0161  &0.2581  & 16.01& -0.9787 \\
 1&0.5&0.0143    &0.3611  &  25.21& -0.9889 \\
 \hline
 2&0&0.3011& 1.3139 &4.66&0 \\
  2&0.3 &0.0266   & 0.2684&10.11&-0.9787  \\
 2&0.5 & 0.0229&0.3896  & 17.02&-0.9927  \\
 \hline
 3&0& 2.009 & 4.3545 & 4.18 &0 \\
  3&0.3& 0.6169  &3.6319& 6.50 &-0.4312\\
 3&0.5&0.1553  & 1.6862 &11.01&-0.4871  \\
 \hline
\end{tabular}
\end{table}

\begin{table}
\caption{Values of different observables in the energy-like squeezed coherent states }\label{table2}
\centering
\begin{tabular}{l|*{6}{| c}}
 z&  $\gamma$&$(\Delta x)^2$& $(\Delta)^2$&T &$Q(z,\gamma)$\\
\hline
 0.1&  0&0.018   &0.2522  & 14.03&0   \\
  0.1&  0.3&0.0111  &0.27  & 24.25&1.2003   \\
 0.1&  0.5&0.0117  &0.4263  &36.232& 1.6842   \\
 \hline
 0.3&  0& 0.0182    & 0.2523 & 13.83&0    \\
  0.3&  0.3 & 0.0112  & 0.2696 & 23.98&1.1864  \\
 0.3& 0.5&0.0118    &0.4243  &35.89& 1.6799   \\
 \hline
 0.6& 0 &0.0186  & 0.2525 & 13.54&0.0001   \\
  0.6& 0.3&0.0114  & 0.2690 & 23.59&1.1413  \\
 0.6& 0.5&0.0119   & 0.4214 &35.39&1.6652   \\
 
 \hline
 1&  0& 0.0192   & 0.2527 &13.16 &0.0003 \\
  1&  0.3  & 0.0116   & 0.2682 &23.09&1.0435   \\
 1& 0.5  &0.0120  &0.4176  & 34.72 &1.6313 \\
 \hline
 2&  0& 0.0207   &0.2533  &12.25& 0.0013  \\
  2&  0.3 & 0.0122   &0.2664  &21.85& 0.7084  \\
 2&  0.5&0.0123   &0.4086  & 33.11&1.4845 \\
 \hline
 3& 0&0.0223   & 0.2540 &11.38&0.003  \\
  3& 0.3&0.0128 & 0.2647 &20.66&0.3909 \\
 3& 0.5 &0.0127  & 0.4005 &31.54&1.2772  \\
 \hline
\end{tabular}
\end{table}

The results (Tables 1 and 2) show that the dispersion in the position $x$ is sensitive to changes in the coherence $z$, the squeezing $\gamma$ and the type  (oscillator or energy-like) of the states.
For the coherent case,  $\gamma=0$, the energy-like states have a smaller dispersion in position compared to  oscillator-like states for each  value of $z$. There is an increase in dispersion when $z$ increases, which is more significant for the oscillator-like states. For example, when $z=0.6$, $(\Delta x)^2=0.0249$ for the oscillator-like states and $(\Delta x)^2=0.0186$ for the energy-like states and, when $z=2$, $(\Delta x)^2=0.3011$ for the oscillator-like states and $(\Delta x)^2=0.0207$ for the energy-like ones.
When  squeezing ($\gamma \neq 0$) is involved, the dispersion, compared to the coherent states for each $z$, is reduced, like for the harmonic oscillator squeezed coherent states. This reduction is more prominent for oscillator-like states,
while the  energy-like states are less sensitive. For example, for $z=0.6$ and $\gamma=0.3$, we get $(\Delta x)^2=0.0138$ for the oscillator-like states and
$(\Delta x)^2=0.0114$ for the energy-like states (compared with the preceding values for $\gamma=0$).
Thus with squeezing, the dispersion in position is reduced
for both types of states, compared to corresponding coherent states.
Overall, the energy-like states are more stable to the changes in $z$ and $\gamma$ and exhibit a consistent good localization in position.

Interestingly, increasing the localization in $x$ will
increase the dispersion  in $p$.  The dispersion in $p$ is larger for energy-like states for each value of $z$ and $\gamma$ so we get similar results as for the harmonic oscillator.

Now, for  $\gamma=0$, the minimal uncertainty, $(\Delta )^2=0.2522 \simeq1/4$, is achieved  for energy-like states in the coherent case
when $z=0.1$. In fact, it is also the case for all energy-like coherent states. Oscillator-like coherent states exhibit larger uncertainty for $z\geq1$.
Increasing the squeezing  $\gamma$
produces a steady increase in  dispersion $(\Delta )^2$ for energy-like squeezed states;
this is  not the case  for oscillator-like squeezed states when $z \geq 1$.
For example, when $z=0.6$, $\gamma=0.3$, the energy-like states have dispersion $(\Delta x)^2=0.0114$ and uncertainty  $(\Delta)^2=0.2690$, the oscillator-like states have dispersion $(\Delta x)^2=0.0138$ and uncertainty $(\Delta)^2=0.2617$. When
  $z=3$,  $\gamma=0.3$, the energy-like states have
good localization $(\Delta x)^2=0.0128$  and  uncertainty $(\Delta)^2=0.2647$,
while for oscillator-like states $(\Delta x)^2=0.6169$ and $(\Delta)^2=3.6319$.

\begin{figure}[h!]
\centering
\includegraphics[width=.7\textwidth]{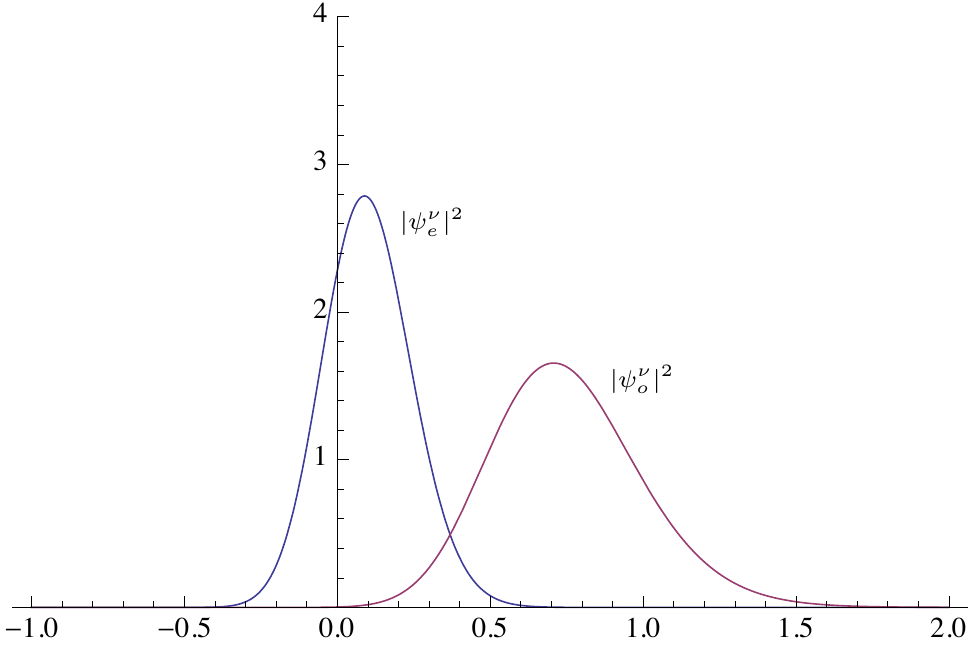}
\caption{Density probability $|\Psi^{\nu}(2, 0,x)|^2$ for Morse
 case in the energy-like (blue) and oscillator-like
(red) coherent states.}
\label{Fig1}
\end{figure}

\begin{figure}[h!]
\centering
\includegraphics[width=.6\textwidth]{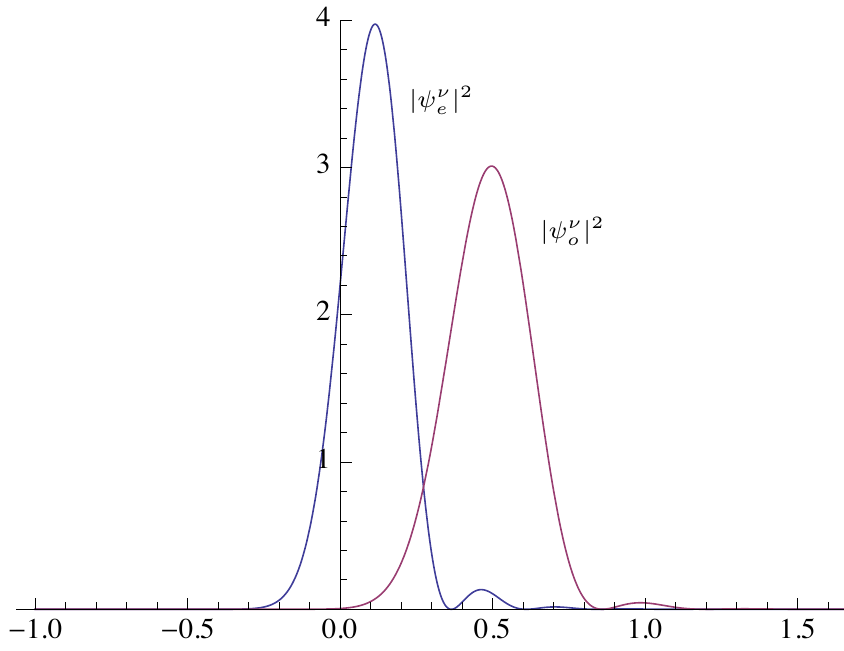}
\caption{Density probability $|\Psi^{\nu}(2, 0.6,x)|^2$ for Morse
 case in the energy-like (blue) and oscillator-like
(red) squeezed coherent states for $z=2$.}
\label{Fig2}
\end{figure}

Thus, if the aim of the construction were to obtain well-localized states in position which also satisfy the minimum uncertainty relation, the
tables demonstrate that the energy-like squeezed coherent states
have a better localization and are more stable. This is
illustrated by the graphs of the comparison of  the density probabilities $|\Psi^{\nu}(z, \gamma,x)|^2$ associated with the energy-like and oscillator-like squeezed coherent states for $(z, \gamma)=(2,0)$ in Fig. \ref{Fig1} and $(z, \gamma)=(2,0.6)$ in Fig.  \ref{Fig2}

The quantum noise $T$ mainly shows that the dispersion in $p$ is in most cases much larger than the one in $x$.
Similarly to the case of the harmonic oscillator, minimal total noise is achieved for given $z$ for  the coherent case ($\gamma=0$) for both types of states.
The squeezing gives rise to an increase of  quantum noise for both types of states.
However, the increase is more significant for energy-like  states, especially when $z\geq1$,
(because the dispersion in $p$ does not increase so quickly for oscillator-like states).
For example, when $z=3$, $\gamma=0.3$, the quantum noise $T=6.50$ for oscillator-like states
while for energy-like states $T=20.66$.
For a given state of  squeezing $\gamma$, the noise decreases with increase in $z$ and less noise is observed  for oscillator-like states.
Thus, if the aim of the construction was to produce less "noisy" states, oscillator-like squeezed coherent states are a better choice, however they show deviation from the minimum uncertainty principle.

Regarding the statistical properties, Table 1 shows  that oscillator-like squeezed states are very similar   to the  states of the harmonic oscillator. For coherent states, $\gamma=0$, Mandel's $Q$-parameter is always zero. For the squeezing in $p$, {\it ie}  if $\gamma <0$,
we have essentially a sub-Poissonian statistics. The case of the squeezing in $x$, {\it ie} if $\gamma>0$, is more interesting since we see that, as in the case of harmonic oscillator,  $Q(z, \gamma)$ exhibits all possible statistics.

    \begin{figure}[h!]
\centering
\includegraphics[width=.65\textwidth]{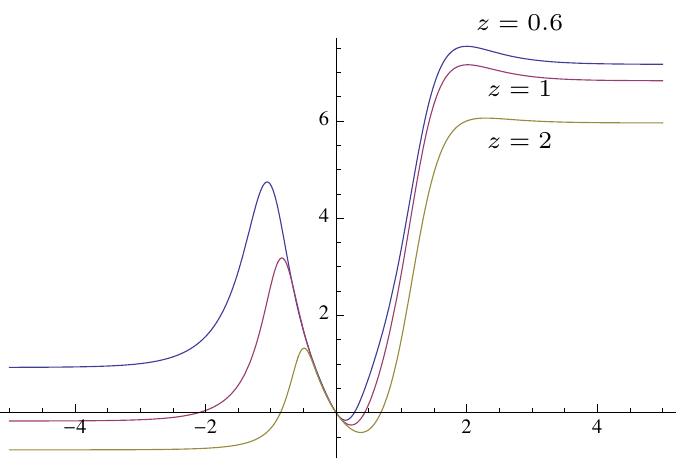}
\caption{ Mandel parameter $Q(z, \gamma)$ for the oscillator-like squeezed states as a function of $r$ such that $\gamma=\tanh r$ for $z= 0.6,1, 2$.}
\label{Fig3}
\end{figure}

\begin{figure}[h!]
\centering
\includegraphics[width=.65\textwidth]{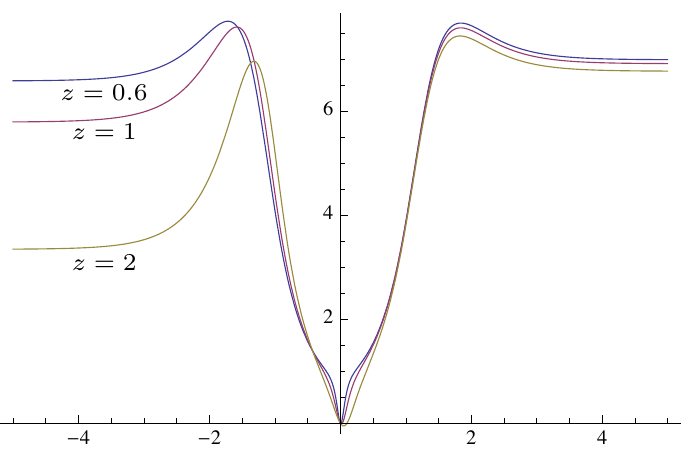}
\caption{Mandel parameter $Q(0, \gamma)$ for the energy-like squeezed states as a function of $r$ such that $\gamma=\tanh r$.}
\label{Fig4}
\end{figure}

Fig. \ref{Fig3} shows the behaviour of Mandel's parameter $Q(z, \gamma)$ for oscillator-like states as a function of $r$ such that $\gamma=\tanh r$  for three cases when  $z=0.6,1, 2$.  The bunching ($Q>0$) is very prominent in these three cases,  has local maxima  when $r>0$ and when $r<0$, it increases  when the values of $z$ decrease. In our example it is the strongest for $z=0.6$. Anti-bunching is less prominent and it is observed in all three cases. The effect is more visible when $z$ increases (see, for $z=2$). A local minimum appears when  $r>0$.

Table \ref{table2} gives the statistical properties of the energy-like states. They exhibit Poissonian statistics when $\gamma=0$, as Mandel's parameter $Q\rightarrow 0$. The bunching effect is clearly observable and is stronger compared to  the oscillator-like states.
Figure 4 shows Mandel's parameter as a function of $r$ for the energy-like states for $z=0.6, 1, 2$. The anti-bunching is weak and appears around $r=0$.

    \begin{figure}[h!]
\centering
\includegraphics[width=.65\textwidth]{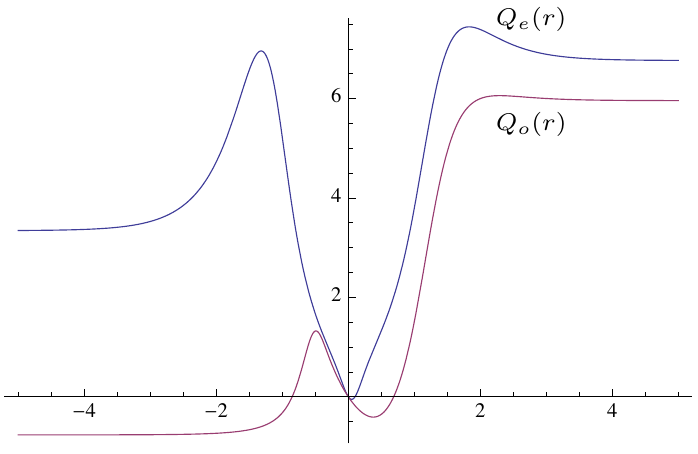}
\caption{ Comparison between the Mandel parameter for the energy-like (blue) and oscillator-like (red) squeezed states as a function of $r$ such that $\gamma=\tanh r$ for $z=2$.}
\label{Fig5}
\end{figure}

\begin{figure}[h!]
\centering
\includegraphics[width=.65\textwidth]{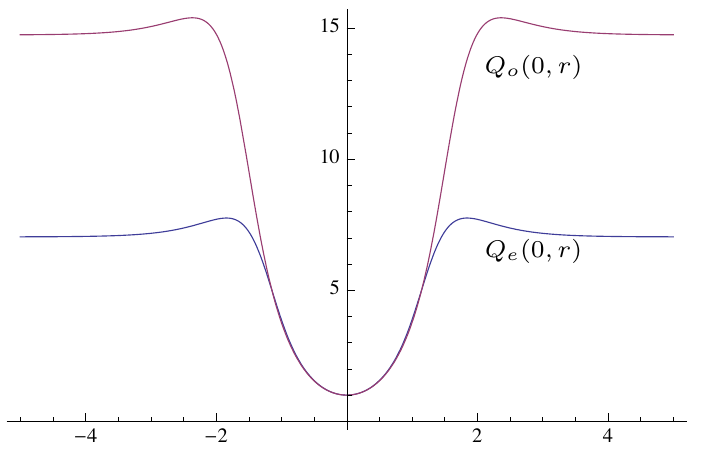}
\caption{Comparison of Mandel parameter $Q(0,\gamma)$ for the squeeze vacuum for the energy-like (blue) and oscillator-like (red) squeezed states as a function of $r$ such that $\gamma=\tanh r$.}
\label{Fig6}
\end{figure}

The comparison between Mandel's parameter for the energy-like and oscillator-like squeezed states  for $z=2$ is given in Fig \ref{Fig5}, there are clear similarities. The bunching  and anti-bunching  are observed  for both types of states,  the bunching is stronger for energy-like states for all values of $r$.  More significant anti-bunching is observed  for oscillator-like states showing a steady effect for  $r<0$ and a minimum for  $r>0$.

The case of squeeze vacuum $z=0$ is interesting. Fig  \ref{Fig6} shows  bunching behaviour of $Q(0, \gamma)>0$ in vacuum for  the energy-like  and oscillator-like squeezed states  in terms of $r$ again. One can clearly observe the symmetry.

\begin{figure}[h!]
\centering
\includegraphics[width=.7\textwidth]{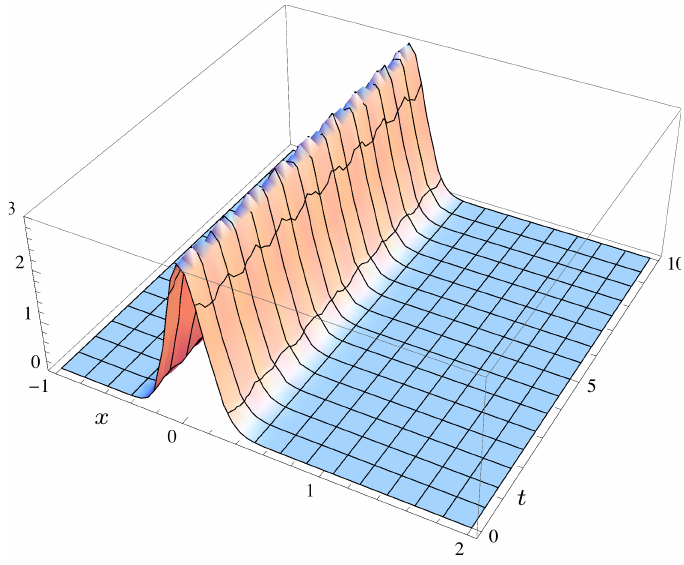}
\caption{ Time evolution of  the density probability $|\Psi^{\nu}_e(0.3, 0,x;t)|^2$ for energy-like coherent states.}
\label{Fig7}
\end{figure}

\begin{figure}[h!]
\centering
\includegraphics[width=.7\textwidth]{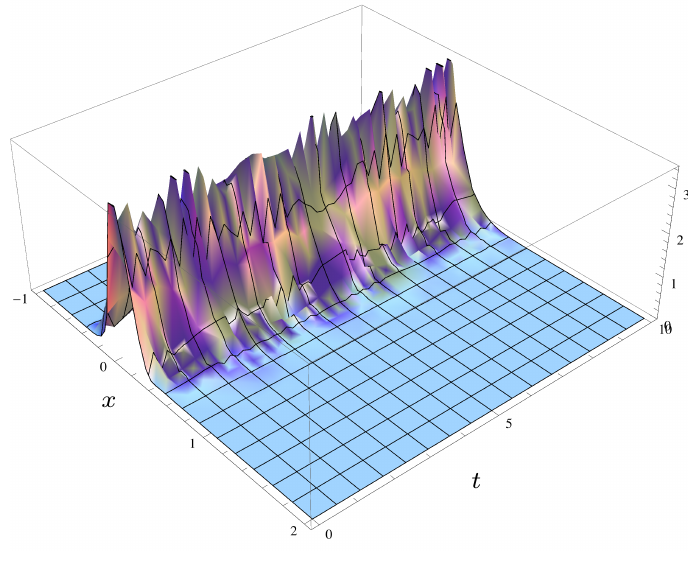}
\caption{ Time evolution of  the density probability $|\Psi^{\nu}_e(0.3, 0.3,x;t)|^2$ for the energy-like squeezed states.}
\label{Fig8}
\end{figure}

Finally, let us mention that time evolution for  our squeezed coherent states is computed as
usual and we get
 \begin{equation}
\Psi^{\nu}(z, \gamma, x;t)=  \frac{1}{\sqrt{{\cal{N}}^{\nu} (z, \gamma)}}
\sum_{n=0}^{[p]-1}  \frac{Z(z,\gamma,n)}{\sqrt{\rho(n)}} e^{- \frac{ i E_n}{\hbar}t} \psi_n^{\nu} (x).
\end{equation}
We thus ask if the squeezed coherent states are stable in time. Fig \ref{Fig7} shows a good stability as time evolves for the density probability $|\Psi^{\nu}_e(0.3, 0,x;t)|^2$ in the energy-like coherent states.
Fig \ref{Fig8} shows the effect of squeezing, $\gamma=0.3$, the states are less stable in time in this case. Similar behaviour is obtained for the oscillator-like states. Let us remind that such conclusions were valid for the harmonic oscillator coherent and squeezed states.

The calculations of observables for other diatomic molecules can be done in a similar way.  For example, we have done such calculations for the molecule $^{133}{\rm Cs}_2$ which has a larger value for $\nu$. Indeed, we have $\nu\approx 524.55$ and thus $[p]=261$. We have obtained similar behaviour for both types of states that is not relevant to produce here.

\section {Conclusions \label{sec:conclusions}}

In this paper, we have reviewed the general results on uncertainty relations and  squeezed coherent states for a quantum system with infinite spectrum and introduced squeezed coherent states of a quantum system with a finite discrete energy spectrum described by the Morse potential.

We have defined two different types of ladder operators, oscillator-like operators that gives rise to a $h(2)$ algebra and energy-like operators associated with a $su(1,1)$ algebra. We have constructed with these operators two types of states of the Morse system:  oscillator-like and energy-like squeezed coherent states.

Even if both types of operators are satisfying algebra structures, the construction of squeezed coherent states has not been realized using the displacement $D$ and squeezed $S$ operators  as introduced in (\ref{dees}) for the harmonic oscillator. Indeed, since we deal with a finite number of eigenstates for the Morse potential, the action of these operators is not well defined even if we take a finite development of the exponentials. Firstly, as we have mentioned in Section 3.2, the last admissible state is not in general canceled when we act on it with $A^+$. Secondly, the Baker-Campbell-Hausdorff formulae for expanding $D$ and $S$ (as products of exponentials of simple operators) are not necessarily valid (see, for example \cite{Sasaki}).

We have considered the  observables of a Morse system, such as dispersion in position, momentum and uncertainty, and investigated the behaviour of our states regarding localization and minimum uncertainty. Note that the calculation of the dispersions and mean values is done
using numerical integration since  $x$ and $p$ are not linear
combination of the annihilation and creation operators in the case
of the Morse potential. It would be interesting to find a way to
analyse the behaviour of these quantities in an analytic way
(dependance in $z$ and $\gamma$).

We have used  the uncertainties  in position and momentum to analyse the quantum noise and Mandel's parameter to investigate the statistical properties of our states.
The oscillator-like squeezed coherent states are closely related to the similar states of the harmonic oscillator. However, these states do not have a very good localization and exhibit a certain deviation from the minimal uncertainty principle. The energy-like squeezed coherent states exhibit a good localization in position and minimize better the uncertainty relation.
They are more stable regarding changes in coherence and squeezing. These states, however, generate stronger quantum noise.  Both types of states exhibit Poissonian statistics in the coherent case, and when squeezing is involved  super- and sub-Poissonian statistics and associated with them effects of bunching and anti-bunching.

\appendix\section{Solution of a three term recurrence relation}

 In this appendix, we are discussing the resolution of the recurrence relation (\ref{recurrel}) which appears when we want to construct  squeezed coherent states as eigenstates of a linear combination of ladder operators. We will take the explicit form of $k(n)$ as given by
 \begin{equation}
k(n)=n(A-n), \ A\in {\mathbb{R}},
\label{knquad}
\end{equation}
so that the recurrence relation becomes
  \begin{equation}
Z(z,\gamma,n+1)- z \ Z(z,\gamma,n) +\gamma\ n(A-n) \ Z(z,\gamma,n-1)=0,  \ n=1,2,...
\label{recurrelk}
\end{equation}
Without restriction we take $Z(z,\gamma,0)=1$ and thus  $Z(z,\gamma,1)=z$. Since we know the solution for the harmonic oscillator (i.e. when $k(n)=n$), we follow the same lines to solve (\ref{recurrelk}). At this stage, we solve the relation (\ref{recurrelk}) for an infinite sequence of values of $n$. 

We introduce the new complex variable $w=\frac{z}{\sqrt{2\gamma}}$ and we take\begin{equation}
Z(z,\gamma,n)=(\frac{\gamma}{2})^{\frac{n}{2}} f(n, w),
\label{bigz}
\end{equation}
We thus get a new recurrence relation on the functions $f(n, w)$:
 \begin{equation}
f(n+1,w)- 2 w \ f(n,w) +2 n(A-n) \ f(n-1,w)=0, \ f(1,w)=2 w, \ f(0, w)=1, \ n=1,2,...
\label{recurrencef}
\end{equation}
It is easy to see that $f(n, w)$ is in fact a polynomial of degree $n$ in $w$. Moreover, it can be expressed in terms of hypergeometric functions of the type ${}_2 F_1$. We explicitly get
\begin{equation}
 f(n,w)= 2^{\frac{n}{2}} (-A+1)_n \  {}_2 F_1\left(\begin{matrix}-n, - {\frac{w}{\sqrt2}}+{\frac{1-A}{2}} \\1 - A\end{matrix};2\right)
\end{equation}
where $(-A+1)_n$ is the usual notation for the Pochhammer symbol
\begin{equation}
(a)_n= a(a+1)(a+2)...(a+n-1)=\frac{\Gamma(a+n)}{\Gamma(a)},
\end{equation}
and the hypergeometric function is in fact a polynomial in $w$ since we have
\begin{equation}
{}_2 F_1\left(\begin{matrix}-n, -v\\1 - A\end{matrix};2\right)= \sum_{k=0}^{n} \frac{2^k}{k!} \frac{(-n)_k(-v)_k}{(-A+1)_k}.
\end{equation}

The original function $Z(z,\gamma,n)$ may thus be written as
\begin{equation}
Z(z,\gamma,n)= \gamma^{\frac{n}{2}} (-A+1)_n \ {}_2 F_1\left(\begin{matrix}-n, - {\frac{z}{2 \sqrt\gamma}}+{\frac{1-A}{2}}  \\ 1 - A\end{matrix};2\right).
\end{equation}
It is valid for any real value of $A$ and in fact, we see that the first polynomials of the sequence are given by
\begin{eqnarray}
Z(z,\gamma,0)&=&1, \ Z(z,\gamma,1)=z, \nonumber \\
Z(z,\gamma,2)&=&z^2-(A-1)\gamma,\nonumber \\
Z(z,\gamma,3)&=&z^3 - (3A-5) \gamma z,\nonumber \\
Z(z,\gamma,4)&=&z^4 - 2 (3A-7) z^2 \gamma + 3 (A-1)(A-3) \gamma^2\nonumber.
\end{eqnarray}

For the special case where $A$ is an integer, we see that the recurrence relation  (\ref{recurrencef}) splits in two different ones. Indeed, we get, first, a finite sequence of $f(n,w)$ satisfying (\ref{recurrencef}) for $n=1,2, ..., A-1$
and, second, an infinite sequence of $f(n,w)$ for $n=A,A+1, ...$ satisfying the recurrence relation
 \begin{equation}
f(A+k+1,w)- 2 w \ f(A+k,w) -2k(A+k) \ f(A+k-1,w)=0, \ k=0,1,2,...
\label{recurrenceinf}
\end{equation}
Since for $k=0$, we get $f(A+1,w)=2w f(A,w)$, we can write $f(A+k,w)=h(k,w) f(A,w)$ where $h(k,w)$ is  a polynomial of degree $k$ in $w$ satisfying the recurrence relation
 \begin{equation}
h(k+1,w)- 2 w \ h(k,w) +2 k(-A-k) \ h(k-1,w)=0, \ h(1,w)=2 w, \ h(0, w)=1, \ k=1,2,...
\label{recurrencek}
\end{equation}
which is (\ref{recurrencef}) where $A$ has been replaced by $-A$. The polynomials $h(k,w)$ are thus given by 
\begin{equation}
 h(k,w)= 2^{\frac{k}{2}} (A+1)_k \  {}_2 F_1\left(\begin{matrix}-k, - {\frac{w}{\sqrt2}}+{\frac{1+A}{2}} \\ 1+A\end{matrix};2\right),  \ k=0,1,2,...
\end{equation}
The solutions  $f(n,w)$ satisfying (\ref{recurrencef}) for $n=1,2, ..., A-1$ are in fact associated to a finite sequence of Krawtchouk polynomials while the solutions  $h(n,w)$ for $n=0,1,...$ are associated with Meixner polynomials \cite{Koe}. Let us mention that they both satisfy discrete orthogonality relations on the variable $w$ but these are not relevant in our context since $w$ is a continuous parameter.

\section*{Acknowledgements}

The authors acknowledge the support of research grants from NSERC of Canada. This
work has been done while V. Hussin visited Northumbria University (as visiting professor and sabbatical leave). This institution is acknowledged for hospitality. The authors thank J. Van der Jeugt for helpful discussions on special functions and orthogonal polynomials.

\end{document}